\RequirePackage{fix-cm}
\documentclass[smallextended]{svjour3}       
\smartqed  
\usepackage{graphicx}

\usepackage[numbers]{natbib}
\usepackage{url} 
\usepackage{hyperref}
\usepackage[dvipsnames]{xcolor}
\usepackage[normalem]{ulem}
\usepackage{amsmath}
\usepackage{amssymb}

\begin{document}

\authorrunning{Pignatari et al.}
\titlerunning{The s process in massive stars and neutron-capture rates}

\title{The s process in massive stars, a benchmark for neutron capture reaction rates}



\author{Marco Pignatari$^{1,2,3,7}$, Roberto Gallino$^{4}$, Rene Reifarth$^{5,6,7}$}



\institute{
$^{1}$Konkoly Observatory, Research Centre for Astronomy and Earth Sciences, HUN-REN, H-1121 Budapest, Konkoly Thege M. \'ut 15-17, Hungary \\ 
           $^{2}$CSFK, MTA Centre of Excellence, Budapest, Konkoly Thege Miklós út 15-17, H-1121, Hungary \\ 
           $^{3}$E.A. Milne Centre for Astrophysics, University of Hull, Hull HU6 7RX, UK \\ 
           $^{4}$Dipartimento di Fisica, Università di Torino, 10125 Torino, Italy\\
          $^{5}$ Goethe University Frankfurt, Max-von-Laue-Strasse 1, Frankfurt am Main 60438, Germany \\           
          $^{6}$ Los Alamos National Laboratory, Los Alamos, NM, 87544, USA \\ 
          $^{7}$ NuGrid Collaboration, \url{http://nugridstars.org} \\           
}

\date{Received: date / Accepted: date}

\maketitle

\begin{abstract}
A clear definition of the contribution from the slow neutron-capture process (s process) to the solar abundances between Fe and the Sr-Zr region is a crucial challenge for nuclear astrophysics. Robust s-process predictions are necessary to disentangle the contribution from other stellar processes producing elements in the same mass region. Nuclear uncertainties are affecting s-process calculations, but most of the needed nuclear input are accessible to present nuclear experiments or they will be in the near future. Neutron-capture rates have a great impact on the s process in massive stars, which is a fundamental source for the solar abundances of the lighter s-process elements heavier than Fe (weak s-process component). In this work we present a new nuclear sensitivity study to explore the impact on the s process in massive stars of 86 neutron-capture rates, including all the reactions between C and Si and between Fe and Zr. We derive the impact of the rates at the end of the He-burning core and at the end of the C-burning shell, where the $^{22}$Ne($\alpha$,n)$^{25}$Mg reaction is is the main neutron source. We confirm the relevance of the light isotopes capturing neutrons in competition with the Fe seeds as a crucial feature of the s process in massive stars. For heavy isotopes we study the propagation of the neutron-capture uncertainties, finding a clear difference of the impact of Fe and Co isotope rates with respect to the rates of heavier stable isotopes. 
The local uncertainty propagation due to the neutron-capture rates at the s-process branching points is also considered, discussing the example of $^{85}$Kr. The complete results of our study for all the 86 neutron-capture rates are available online. Finally, we present the impact on the weak s process of the neutron-capture rates included in the new ASTRAL library (v0.2).  
\end{abstract}
%
%
\section{Introduction}
\label{sec:intro}

The slow-neutron capture process (s process) in massive stars (M$\gtrsim$10M$_{\odot}$) is the main producer of the weak s-process component in the Solar System, accounting for the bulk of the s-process abundances between Fe and Sr \cite{kaeppeler:11}. This includes most of the solar amount of Cu, Ga and Ge, and a significant contribution to the isotopic abundance pattern of other elements in this mass region \cite{raiteri:91, raiteri:91b, the:07, pignatari:10}. 
Between Fe and Sr a zoo of other nucleosynthesis processes have been identified as potential stellar sources relevant for galactic chemical evolution (GCE). Among them are the neutrino-driven wind ejecta from core-collapse supernovae (CCSNe) \cite{froehlich:06,farouqi09,roberts:10,arcones:11,wanajo:18}, electron-capture supernovae \cite{wanajo11,jones:19} and the intermediate neutron capture process (i-process) \cite{roederer:16,cote:18,banerjee:18}. 
While it is commonly assumed that the solar weak s-process component is made by massive stars alone, its whole production cannot be uniquely attributed to a single astrophysical source. The weak s process should be considered instead as the direct product of the activation of the $^{22}$Ne($\alpha$,n)$^{25}$Mg reaction as a source of free neutrons in different types of stars. This is the case for massive stars, for massive Asymptotic Giant Branch (AGB) stars (4 M$_\odot$ $\gtrsim$ M $\gtrsim$ 7 M$_\odot$) and possibly super AGB stars (7 M$_\odot$ $\gtrsim$ M $\gtrsim$ 9 M$_\odot$) \cite{karakas14dawes, jones14}. At present, the GCE contribution of massive AGB stars to the s-process isotopes is not well separated from low-mass AGB stars, which power the production of the main s-process component elements via the activation of the $^{13}$C($\alpha$,n)$^{16}$O \cite{travaglio:04,prantzos:20,kobayashi:20}. The effective contribution of super AGB stars is still highly uncertain, and it is not even clear if super AGB stars are relevant sources of heavy elements for GCE, or if they are s-process sources, or i-process sources (with the activation of the $^{13}$C($\alpha$,n)$^{16}$O reaction) or both \cite{doherty:14,jones:16}.

Massive stars were the first astrophysical source of the s process identified in stellar simulations, with the nuclear reaction $^{22}$Ne($\alpha$,n)$^{25}$Mg as the main neutron source \cite{peters:68}. It is activated in He-burning and C-burning hydrostatic conditions, where $^{22}$Ne is made starting from initial abundances of C, N and O by chains of nuclear reactions \cite{couch:74, busso:85, langer:89, prantzos:90, raiteri:91, raiteri:91b, the:07, pignatari:10}. The final CCSNe explosion can still modify some of the s-process abundances ejected, depending on, e.g., the explosion energy and the amount of $^{22}$Ne left in the convective C-shell ashes \cite{rauscher:02, pignatari:10, pignatari:16,sukhbold:16}.    

The origin of the $^{22}$Ne neutron source and the nature of the s-process seed $^{56}$Fe are 
responsible for the fact that s-process production is secondary in massive stars  
\cite{prantzos:90, baraffe:92, raiteri:92}. Such a paradigm can be changed in fast-rotating massive stars at low metallicity. In fast rotators, primary $^{22}$Ne can be made during He-burning independently from the initial composition, which boosts the s-process production by orders of magnitude compared to non-rotating massive stars \cite{pignatari:08, frischknecht:16, limongi:18, choplin:18}. While it is still matter of debate if 
such a contribution is relevant for the formation of the weak s-process component in the Solar System \cite{pignatari:08,prantzos:20}, several works have explored the signature of its contribution in stellar and in galactic archaeology at low metallicities \cite{chiappini:11, cescutti:13, yong:17, barbuy:21}. The termination point of the s-process production in fast-rotating massive stars is also uncertain, with the potential to reach high yields up to the Ba s-process peak \cite{frischknecht:16, limongi:18}. At the moment, the large impact of nuclear uncertainties of, e.g, the $^{22}$Ne($\alpha$,n)$^{25}$Mg \cite{pignatari:08} and the $^{17}$O+$\alpha$ reaction rates \cite{best:13, frost-schenk:22} are preventing to constrain the s-process production efficiency of fast-rotating massive stars.    


More in general, nuclear uncertainties have a strong impact on s-process abundance yields in massive star models. And this is true for all metallicities. The main neutron source $^{22}$Ne($\alpha$,n)$^{25}$Mg is mostly competing with the $^{22}$Ne($\alpha$,$\gamma$)$^{26}$Mg reaction during core He-burning, and with the $^{22}$Ne(p,$\gamma$)$^{23}$Na in C-burning conditions \cite[][]{pignatari:10}. 
While the $^{22}$Ne(p,$\gamma$)$^{23}$Na is now well measured for the relevant temperatures \citep[][]{depalo:16}, the impact of the uncertainties in the $\alpha$-capture channels of $^{22}$Ne affects the calculated production of s-process elements by up to a factor of 10 \citep[][]{talwar:16, ota:21}, and it is still matter of debate in the literature \cite{adsley:21, wiescher:23}. Future nuclear experiments in underground facilities will drastically improve the present knowledge of such crucial rates \cite{rapagnani:22}. 

Neutron-capture cross sections relevant for the s process in massive stars need to be also measured with a precision of the order of 20$\%$ or less. The overall impact of neutron-capture rate uncertainties on the weak s process was estimated to be in the order of 2-3 \cite{pignatari:10}. By using a Monte Carlo approach an impact varying between 30$\%$ and a factor of 3 was obtained depending on the s-process isotope, but without considering the impact of the neutron-capture rate uncertainties of light nuclei \cite{nishimura:17}. 

The impact of new neutron-capture cross sections measured in the past decade on the s-process predictions has not been clearly evaluated yet.
More in general, without having a robust set of nuclear reactions it is not possible to fully define the sources of discrepancy between different stellar simulations where the s-process efficiency varies up to a factor of three \citep[see e.g.,][]{sukhbold:16,ritter:18}. This will become a necessary step to improve the robustness of s-process calculations once the new $^{22}$Ne+$\alpha$ rates will be available in the next few years.

The paper is organized as follows. In Section~\ref{sec:methodology} we describe the models used for the simulations, in Section~\ref{sec:nuc results} we present the results obtained for the sensitivity study varying the neutron capture rates relevant for the weak s process. Final considerations and conclusions are given in Section~\ref{sec:conclusions}.

\section{Methodology and nucleosynthesis simulations}
\label{sec:methodology}

We use a trajectory representative of the relevant s-process conditions in massive stars derived from a 25 M$_{\odot}$ star model and solar metallicity \cite{hirschi:04}, and already used for previous nucleosynthesis studies \cite{talwar:16, nishimura:17,roederer:22}. This trajectory is also provided by the CheTEC-INFRA platform $ORCHESTRA$\footnote{\url{http://chanureps.chetec-infra.eu/}} in Zenodo\footnote{\url{https://doi.org/10.5281/zenodo.7852263}}. The available temperature and density evolve from central H-burning up to the end of shell C-burning \cite{nishimura:17}. 
The nucleosynthesis is followed with the post-processing NuGrid code $PPN$ \cite{pignatari:12}. For the main analysis presented in Section \ref{sec:nuc results} we use the current $PPN$ default configuration for the network of nuclear reaction rates \cite{lawson:22}, which is quite similar to the nuclear rates adopted for the previous s-process impact study made using this same trajectory \cite{nishimura:17}. In particular, the neutron capture rates relevant for the s process are mostly provided by KADoNIS v0.3 \cite{dillmann:06}, with a number of more recent updates \cite{massimi:12, lederer:13, heil:14, lederer:14, lugaro:14}. 

In Table \ref{tab: relevant rates} we report the most efficient rates activated over the complete trajectory, excluding the reactions relevant for energy generation (e.g., the 3$\alpha$ and the $^{12}$C+$^{12}$C reactions). The nucleosynthesis fluxes generated by the $\alpha$-capture and neutron-capture rate are activated in both the He core and the C shell, while proton-captures dominate during the C-shell phase. Among the strongest $\alpha$-capture rates we find the $^{22}$Ne($\alpha$,n)$^{25}$Mg neutron source. The strongest proton-capture is the $^{12}$C(p,$\gamma$)$^{13}$N. However, at typical C-shell temperatures the $^{12}$C+p channel is already in balance with the reverse reaction $^{13}$N($\gamma$,p)$^{12}$C, and the net impact on the proton budget shaping the nucleosynthesis is negligible \cite{chieffi:98, pignatari:10}. Both in the He core and in the C shell most of the neutrons produced by the $^{22}$Ne($\alpha$,n)$^{25}$Mg are captured by light isotopes (which are also called neutron "poisons" within the s-process context). This is also confirmed by the neutron-capture reactions listed in Table \ref{tab: relevant rates}. However, some of the neutrons captured by these species are recycled by ($\alpha$,n) reactions: the neutrons captured by $^{20}$Ne are partially re-emitted by the $^{21}$Ne($\alpha$,n)$^{24}$Mg; the $^{17}$O($\alpha$,n)$^{20}$Ne and the $^{17}$O($\alpha$,$\gamma$)$^{21}$Ne$^{21}$Ne($\alpha$,n)$^{24}$Mg chain recycle the neutrons captured by $^{16}$O; and the $^{13}$C($\alpha$,n)$^{16}$O recycle the neutrons captured by $^{12}$C.  

\begin{table}
\centering
\caption{The $\alpha$-capture, proton-capture, and neutron-capture rates with the four relative highest nucleosynthesis fluxes integrated over the full trajectory. }
\label{tab: relevant rates}
\begin{tabular}{ccc}
\hline\noalign{\smallskip}
 $\alpha$-captures   & p-captures   &  n-captures   \\
\noalign{\smallskip}\hline\noalign{\smallskip}
$^{16}$O($\alpha$,$\gamma$)$^{20}$Ne & $^{12}$C(p,$\gamma$)$^{13}$N & $^{20}$Ne(n,$\gamma$)$^{21}$Ne \\
$^{20}$Ne($\alpha$,$\gamma$)$^{24}$Mg & $^{23}$Na(p,$\alpha$)$^{20}$Ne & $^{16}$O(n,$\gamma$)$^{17}$O \\
$^{22}$Ne($\alpha$,n)$^{25}$Mg & $^{27}$Al(p,$\alpha$)$^{24}$Mg & $^{24}$Mg(n,$\gamma$)$^{25}$Mg \\
$^{23}$Na($\alpha$,p)$^{26}$Mg & $^{26}$Mg(p,$\gamma$)$^{27}$Al & $^{12}$C(n,$\gamma$)$^{13}$C \\
\noalign{\smallskip}\hline
\end{tabular}
\end{table}

Starting from the template simulation using the default network described above, we performed a sensitivity study focused on the (n,$\gamma$) rates. This includes a total of 172 nucleosynthesis models, where for each run one of 86 different neutron-capture rate is multiplied or divided by a factor of 1.3. We have considered for the analysis all the neutron-capture rates of the stable isotopes between C and Si and of the stable isotopes between Fe and Zr. In this same region we also considered 20 unstable species located along the s-process path. Several of them are branching-points of the s process, opened by the competition between neutron captures and $\beta$-decays (e.g., $^{63}$Ni and $^{85}$Kr) \cite{lederer:14, bisterzo:15}. 
The factor of 1.3 was chosen for these tests by considering a realistic variation between different experimental (n,$\gamma$) rates for the same reaction. Many of these experimental rates have a reported 1$\sigma$ error smaller than 30$\%$. However, larger variations have been found with newer results compared to existing data (see for instance for the cross sections of Ni \cite{lederer:14} and Cu \cite{heil:08}), and large discrepancies still exist between different experiments (see the $^{30}$Si(n,$\gamma$)$^{31}$Si by \cite{guber:03} and \cite{beer:02}, differing by more than a factor of two). Based on these considerations, the same variation factor was recommended by Franz K\"{a}ppeler twenty years ago to perform a study similar to that presented here. That original work remained unpublished, but it was of great use to motivate future experiments together with Franz and his research group.

\begin{figure}
\centering
  \includegraphics[width=5.85 cm]{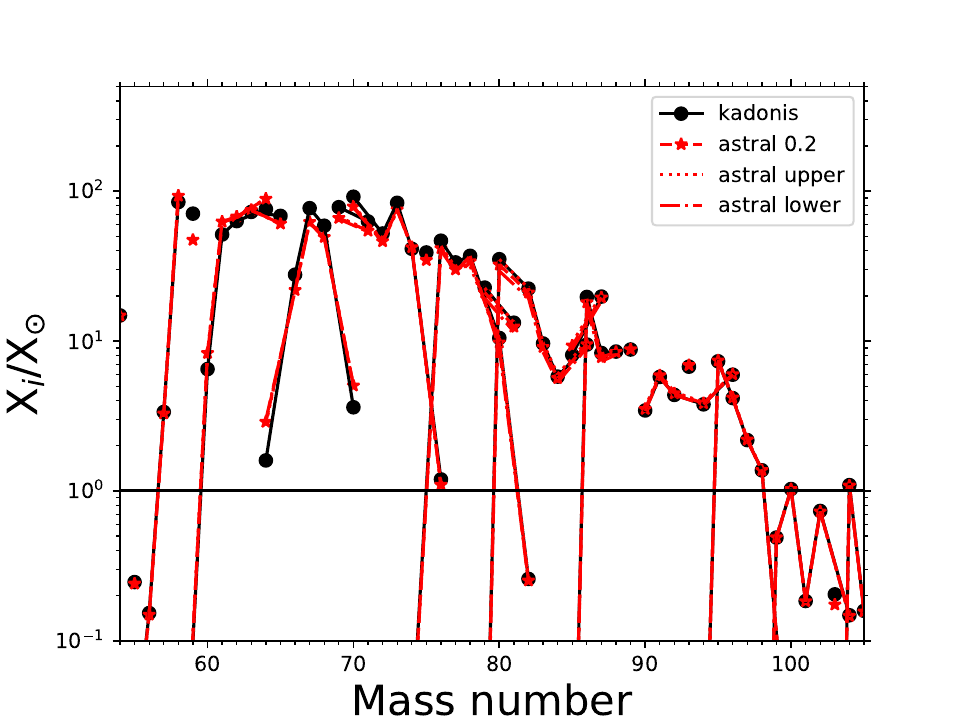}
  \includegraphics[width=5.85 cm]{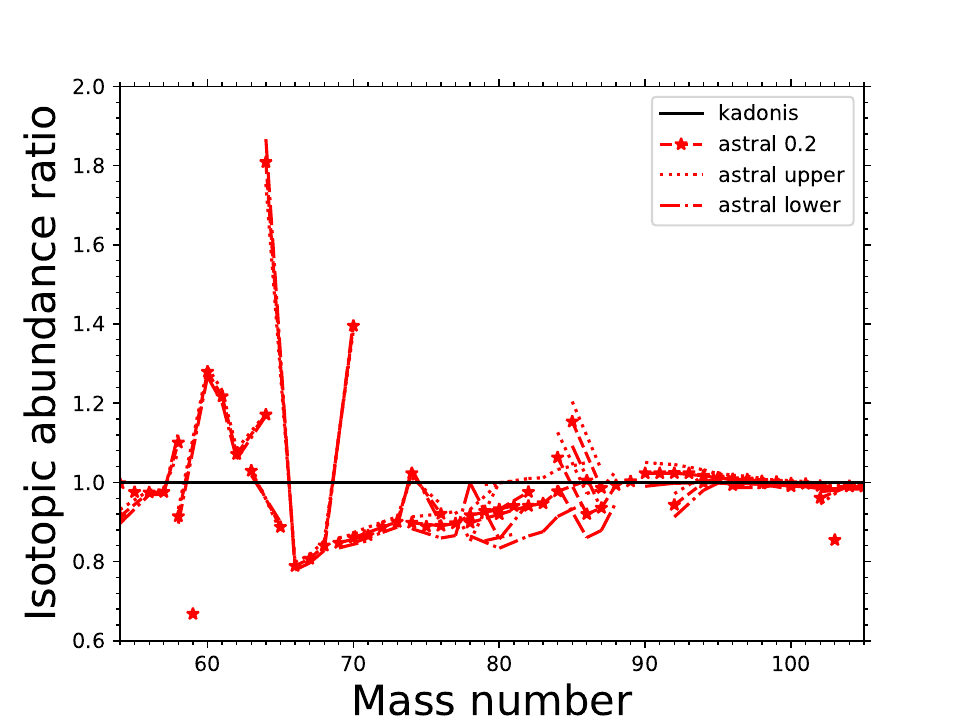}
  \includegraphics[width=5.85 cm]{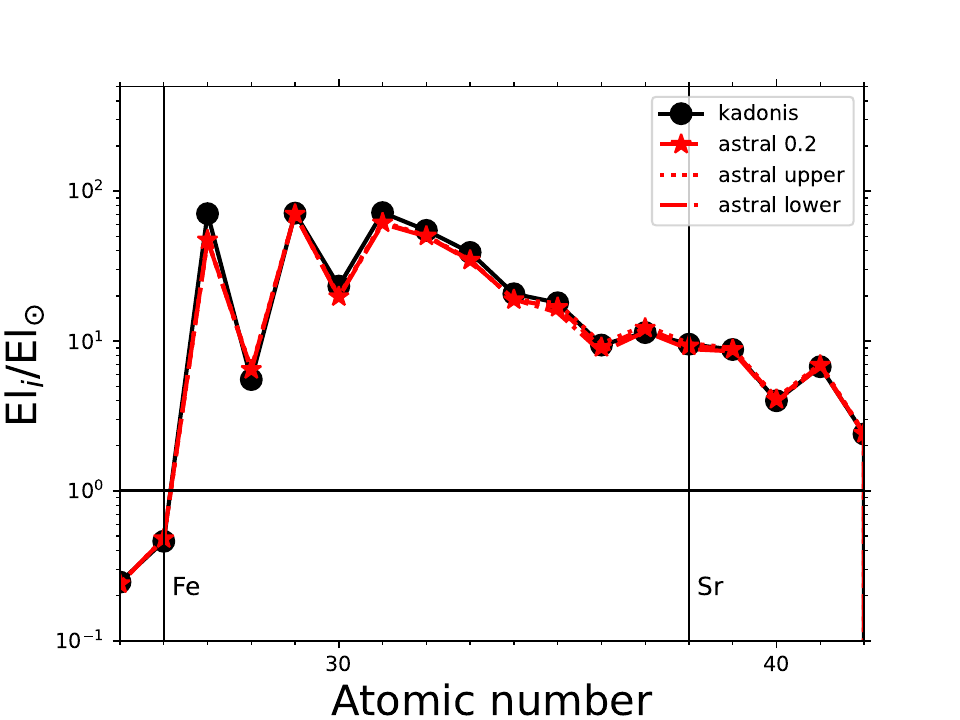}
  \includegraphics[width=5.85 cm]{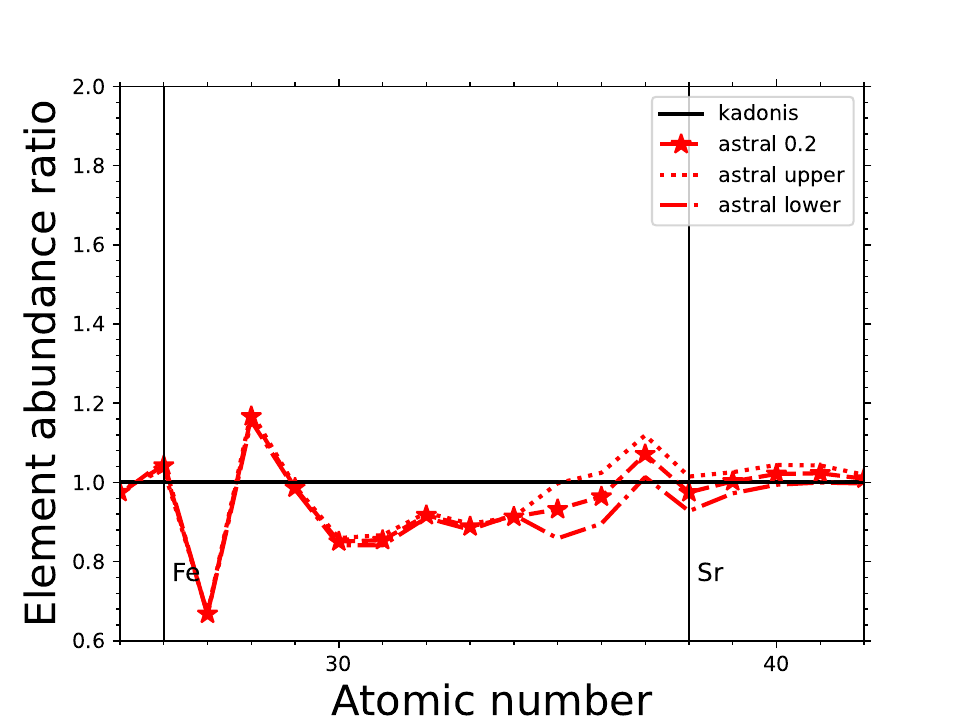}
\caption{The weak s-process abundances at the end of the C shell for the isotopes (upper panels) and the elements (lower panels) between Fe and Nb. The results obtained using the default $PPN$ model are compared with the results obtained using the recommended rates, upper limits and lower limits provided by the ASTRAL library. The production factors are provided on the left panels, while the corresponding abundance ratios (relative to the default model) 
are shown in the right panels.  }
\label{fig:astral}       
\end{figure}

The models introduced here and discussed again in the next section are calculated using the (n,$\gamma$) reactions from the KADoNIS library and the mentioned updates. A new library for neutron-capture cross sections called ASTRAL\footnote{\url{https://exp-astro.de/astral/}} is currently under development \cite{vescovi:23}. The ASTRAL v0.2 includes 122 new (n,$\gamma$) rates with error estimations for all energies. While it is not complete yet, it is interesting to check the impact of the new reaction rates reported in ASTRAL \cite{vescovi:23}. In Figure \ref{fig:astral}, our default model abundances are compared with three sets of abundances calculated using the ASTRAL recommended rates, and their corresponding upper limits and lower limits, when available. The isotopic pattern show several differences, in particular in the Fe-Zn region where the new (n,$\gamma$) rates produce some decreases down to a factor of 1.5 ($^{59}$Co) and some increases up to a factor of 1.8 ($^{64}$Zn). Using the new rates, the elemental s-process production shows a $\sim$30$\%$ lower Co production, an increase of the order of 20$\%$ of Ni, and a decrease of the order of 10-20$\%$ propagated up to Se. Beyond Se the nuclear uncertainties reported by ASTRAL have a larger impact on the s-process abundances, 
making them consistent, within their variations, with the calculations using the default network (Figure \ref{fig:astral}, lower panels). In Figure \ref{fig:astral}, upper panels, the isotopic patterns confirm the overall elemental trends, but the isotope-to-isotope variations are more complex and these features are crucial to take into account for the GCE of s-process isotopes and the reproduction of the s-process isotopic solar composition in the mass region between Fe and Zr \cite{travaglio:04,bisterzo:14,prantzos:20}.

\section{Results of the Nuclear sensitivity study}
\label{sec:nuc results}

A summary of the results is presented in Table \ref{tab:n_sum}, which lists the twenty (n,$\gamma$) rates with the largest impact on the s-process abundances at the end of the He core and at the end of the C shell. The "impact" values listed in the Table are given by the sum of all the differences between the two sets of isotopic production factors in the models, calculated using the same (n,$\gamma$) rate multiplied and divided by a factor of 1.3. Therefore, they
measure the total amount of variations generated on the nucleosynthetic abundances by the variation of the reaction rates. These calculated impact values highlight the impact of the uncertainties having the most extensive propagation effect on the isotopic abundances with the highest production factors. Such an effect is well known to be relevant for the weak s process, involving both isotopes beyond iron (see for instance the famous case of the $^{62}$Ni(n,$\gamma$)$^{63}$Ni, e.g., \cite{nassar:05}) and light isotopes (see, e.g, $^{25}$Mg(n,$\gamma$)$^{26}$Mg \cite{raiteri:91, baraffe:92, massimi:12} and $^{22}$Ne(n,$\gamma$)$^{23}$Ne \cite{busso:85, heil:14}).  
Note that, however, the impact values adopted here should not be considered as the only diagnostic to measure the relevance of (n,$\gamma$) rates. For instance, as we will see later in this section, a more localized impact on the yields with a less extensive propagation is also obtained for unstable isotopes at the s-process branching points.  

All the figures for the 86 reaction rates tested in this analysis are provided online on Zenodo\footnote{\url{https://doi.org/10.5281/zenodo.10124711}}, with the abundances shown at the end of the He core and at the end of the C shell. 

\begin{table}
\centering
\caption{The neutron-capture rates with the largest impact on the abundance distribution are reported at the end of the He core 
and at end of the C shell. 
The impact values (Column 1 and Column 3 for the He core and the C shell, respectively) are defined in the text. 
}
\label{tab:n_sum}
\begin{tabular}{cccc}
\hline\noalign{\smallskip}
 \multicolumn{2}{c}{End of He-core burning}   & \multicolumn{2}{c}{End of C-shell burning}   \\
 \hline\noalign{\smallskip}
 Impact & Reaction & Impact & Reaction \\
\noalign{\smallskip}\hline\noalign{\smallskip}
318.71  &  1FE56+1NEUT$\longrightarrow$1FE57+$\gamma$ & 477.89  &  1FE56+1NEUT$\longrightarrow$1FE57+$\gamma$ \\ 
160.04  &  1MG25+1NEUT$\longrightarrow$1MG26+$\gamma$ & 365.37  &  1MG25+1NEUT$\longrightarrow$1MG26+$\gamma$ \\
146.36  &  1FE58+1NEUT$\longrightarrow$1FE59+$\gamma$ & 307.50  &  1MG24+1NEUT$\longrightarrow$1MG25+$\gamma$ \\
 95.66  &  1FE57+1NEUT$\longrightarrow$1FE58+$\gamma$ & 305.04  &  1NE20+1NEUT$\longrightarrow$1NE21+$\gamma$ \\
 81.60  &  1NI60+1NEUT$\longrightarrow$1NI61+$\gamma$ & 275.19  &  1NI62+1NEUT$\longrightarrow$1NI63+$\gamma$ \\
 78.55  &  1NI62+1NEUT$\longrightarrow$1NI63+$\gamma$ & 244.31  &  1FE58+1NEUT$\longrightarrow$1FE59+$\gamma$ \\
 71.55  &  1MG24+1NEUT$\longrightarrow$1MG25+$\gamma$ & 240.40  &  1ZN68+1NEUT$\longrightarrow$1ZN69+$\gamma$ \\
 65.84  &  1CO59+1NEUT$\longrightarrow$1CO60+$\gamma$ & 229.14  &  1O 16+1NEUT$\longrightarrow$1O 17+$\gamma$ \\
 55.22  &  1CU63+1NEUT$\longrightarrow$1CU64+$\gamma$ & 208.98  &  1CU65+1NEUT$\longrightarrow$1CU66+$\gamma$ \\
 55.18  &  1NI61+1NEUT$\longrightarrow$1NI62+$\gamma$ & 202.74  &  1NI60+1NEUT$\longrightarrow$1NI61+$\gamma$ \\
 54.03  &  1CU65+1NEUT$\longrightarrow$1CU66+$\gamma$ & 184.08  &  1ZN66+1NEUT$\longrightarrow$1ZN67+$\gamma$ \\
 42.39  &  1ZN68+1NEUT$\longrightarrow$1ZN69+$\gamma$ & 137.32  &  1NI61+1NEUT$\longrightarrow$1NI62+$\gamma$ \\
 38.94  &  1ZN66+1NEUT$\longrightarrow$1ZN67+$\gamma$ & 134.21  &  1CO59+1NEUT$\longrightarrow$1CO60+$\gamma$ \\
 37.99  &  1O 16+1NEUT$\longrightarrow$1O 17+$\gamma$ & 127.89  &  1FE57+1NEUT$\longrightarrow$1FE58+$\gamma$ \\
 35.84  &  1SI28+1NEUT$\longrightarrow$1SI29+$\gamma$ & 127.42  &  1CU63+1NEUT$\longrightarrow$1CU64+$\gamma$ \\
 35.25  &  1NE22+1NEUT$\longrightarrow$1NE23+$\gamma$ & 117.48  &  1NA23+1NEUT$\longrightarrow$1NA24+$\gamma$ \\
 34.41  &  1ZN64+1NEUT$\longrightarrow$1ZN65+$\gamma$ & 109.94  &  1ZN67+1NEUT$\longrightarrow$1ZN68+$\gamma$ \\
 21.19  &  1NI58+1NEUT$\longrightarrow$1NI59+$\gamma$ & 109.49  &  1GE70+1NEUT$\longrightarrow$1GE71+$\gamma$ \\
 19.47  &  1ZN67+1NEUT$\longrightarrow$1ZN68+$\gamma$ &  92.94  &  1GA69+1NEUT$\longrightarrow$1GA70+$\gamma$ \\
 19.38  &  1GE70+1NEUT$\longrightarrow$1GE71+$\gamma$ &  92.59  &  1ZN64+1NEUT$\longrightarrow$1ZN65+$\gamma$ \\
\noalign{\smallskip}\hline
\end{tabular}
\end{table}

From the list reported in Table \ref{tab:n_sum}, as expected the neutron-capture rate of the main s-process seed $^{56}$Fe has the largest impact. The impact of light neutron poisons become more relevant in the C shell than in the He-burning phase. The reason is that during C-burning the poisons $^{20}$Ne and $^{23}$Na are directly produced by C fusion reactions, and $^{24}$Mg is mostly made by the $^{20}$Ne($\alpha$,$\gamma$)$^{24}$Mg. Their higher abundances compared to He-burning stellar layers make them more efficient in capturing the neutrons made by $^{22}$Ne \cite{pignatari:10}.

\begin{figure}
\centering
  \includegraphics[width=5.85 cm]{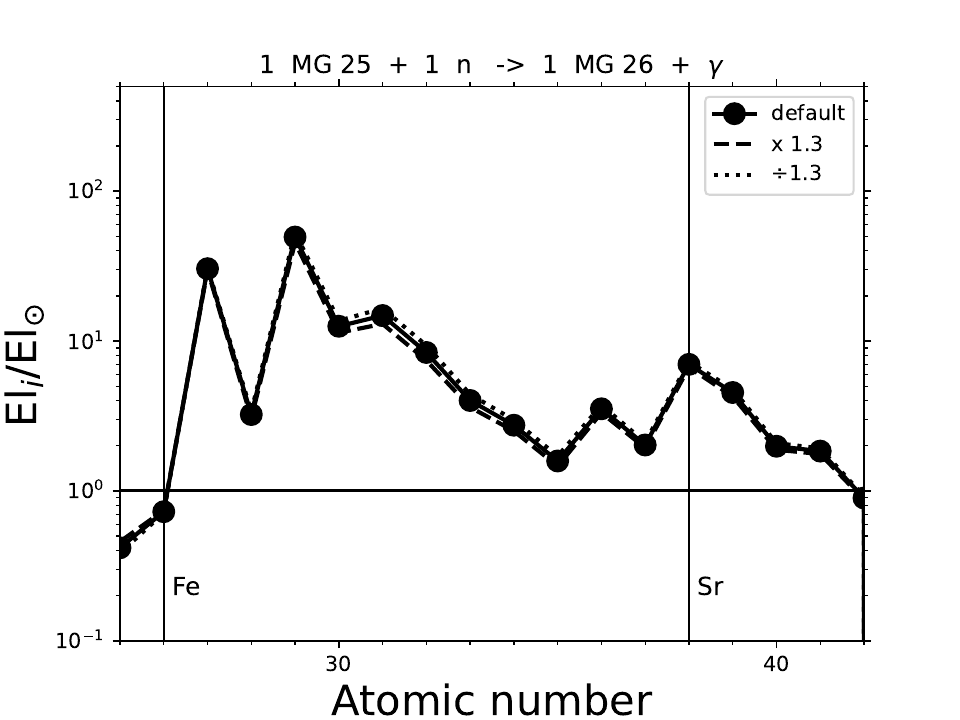}
  \includegraphics[width=5.85 cm]{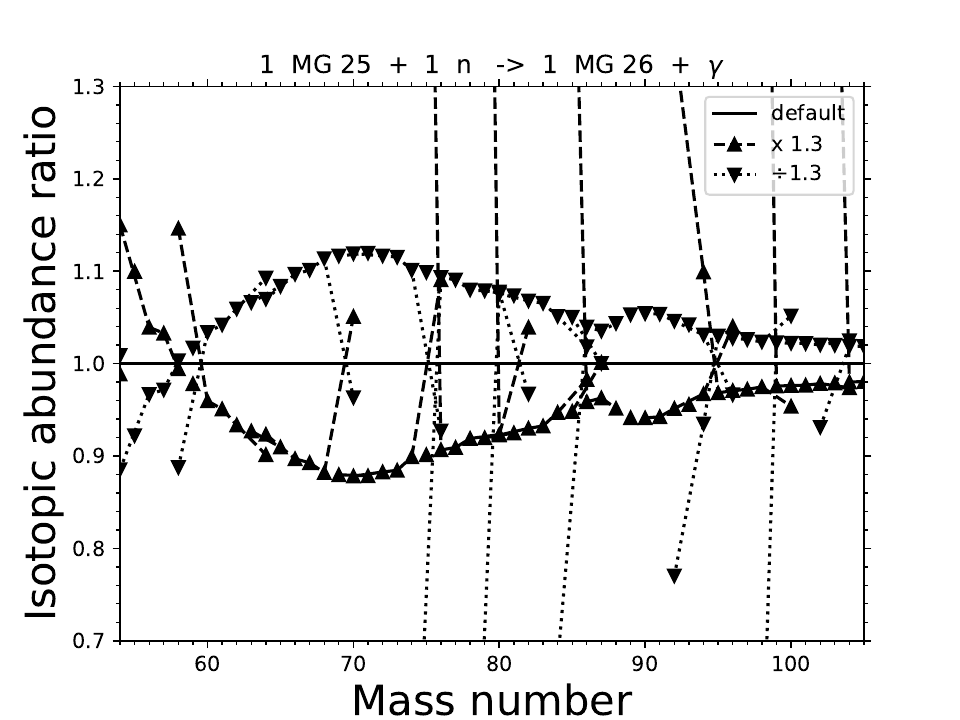}
  \includegraphics[width=5.85 cm]{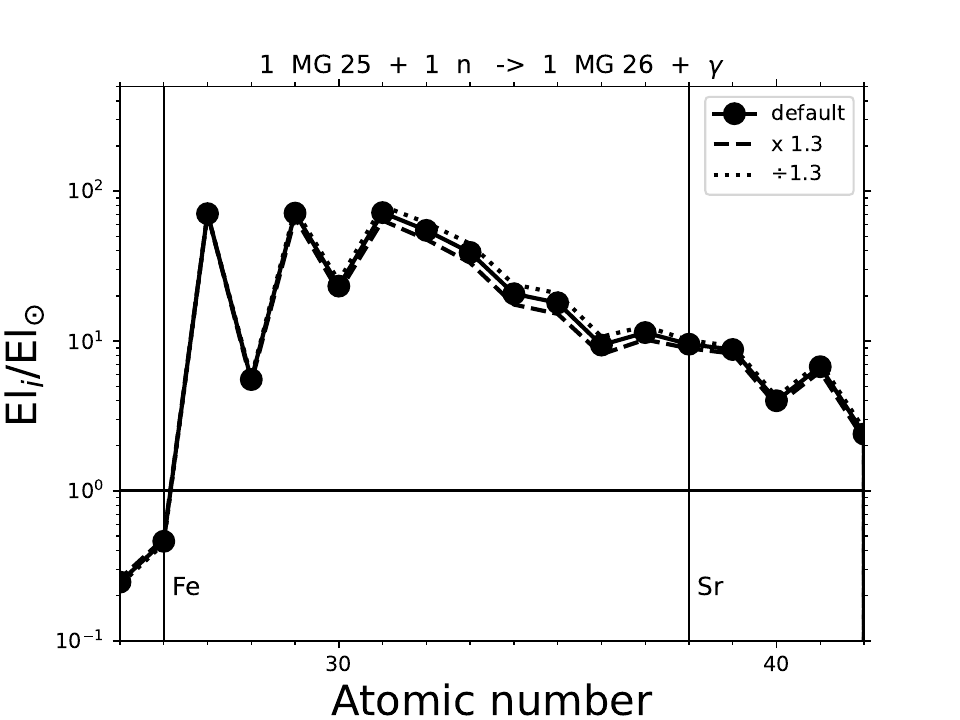}
  \includegraphics[width=5.85 cm]{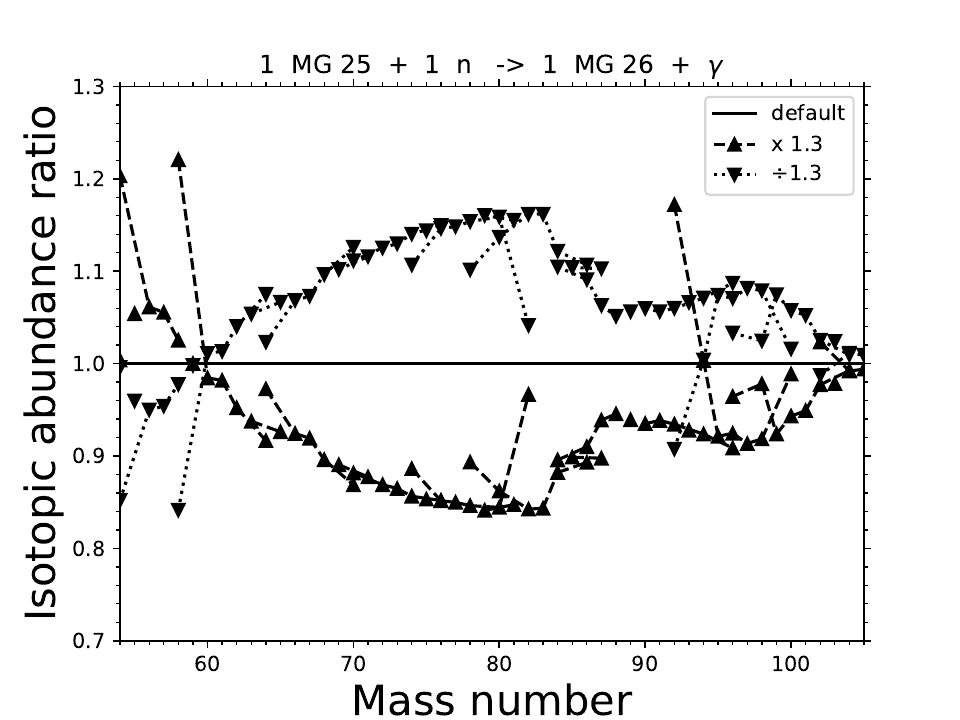}
\caption{The weak s-process abundances at the end of the He core (upper panels) and at the end of the C shell (lower panels). The results obtained using the default model are compared with the same model but multiplying and dividing the $^{25}$Mg(n,$\gamma$)$^{26}$Mg by a factor of 1.3. We report the elemental abundances (left panels) and the isotopic abundance ratios of the test cases normalized to the abundances obtained with the default model.  }
\label{fig:mg25}       
\end{figure}

The neutron poison showing the largest impact for both He-core and C-shell conditions is $^{25}$Mg. In Figure \ref{fig:mg25}, 
we show the results obtained when varying its (n,$\gamma$) cross section. 
The s-process element patterns are not modified overall since the considered variation factor is only of 30$\%$. However, the isotopic pattern is significantly affected. The variations beyond the y-axis scale at the end of the He-core are due to the proton-rich nuclei ($^{74}$Se, $^{78}$Kr, $^{84}$Sr, $^{92}$Mo and $^{98}$Ru). These isotopes are made by the p-process in supernovae \cite{rauscher:13, pignatari:16, roberti:23}, and they are destroyed here by the s process. Therefore, those variations are not relevant for their stellar production. In the Figure \ref{fig:mg25}, right panel, the variation of the $^{25}$Mg neutron-capture rate is propagated over all the s-process isotopes. The impact is however not equal within the abundance distribution: while the Cu-Ga mass region just above Fe is marginally affected, the abundances in the heavier mass region close to the Sr neutron-magic peak shows an inverse change that is quite close to the variation factor applied to the $^{25}$Mg(n,$\gamma$)$^{26}$Mg rate \cite{pignatari:10}.         

\begin{figure}
\centering
  \includegraphics[width=9 cm]{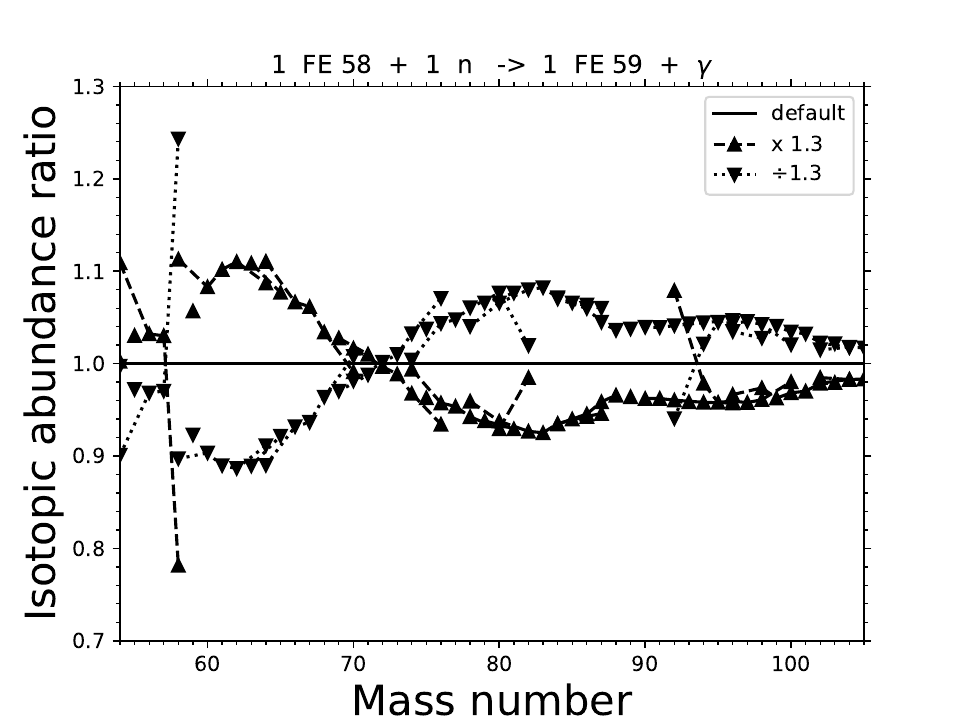}
  \includegraphics[width=9 cm]{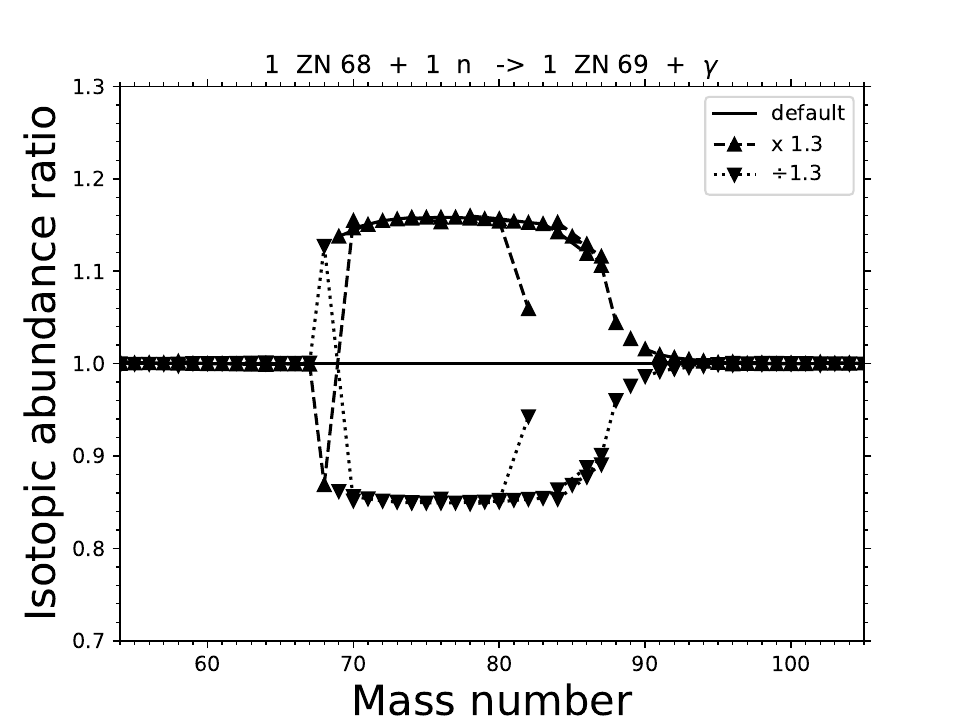}
  \includegraphics[width=9 cm]{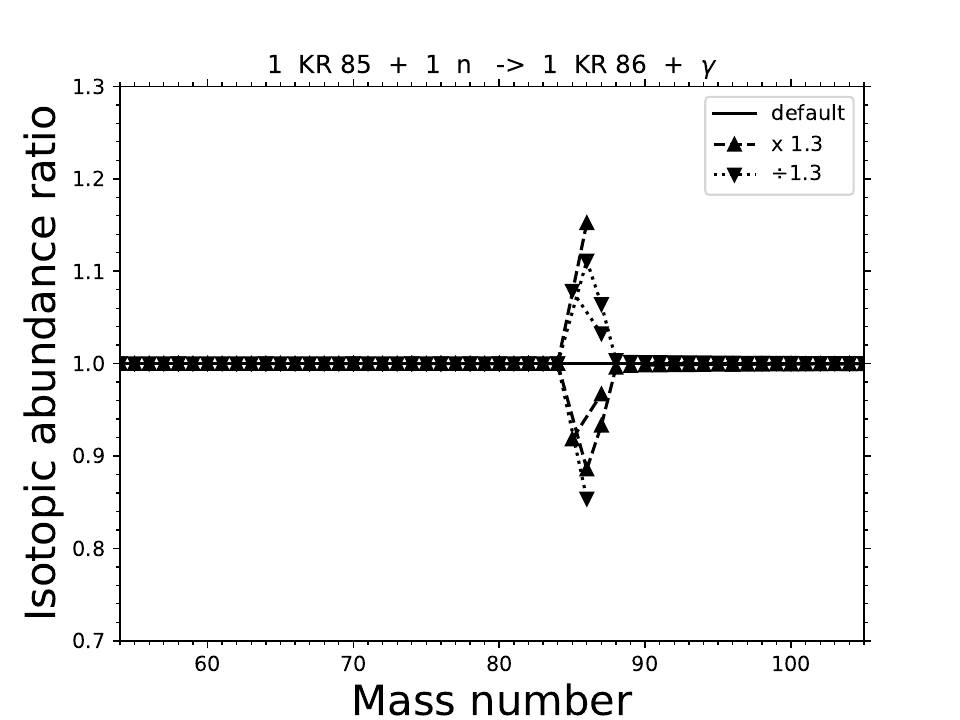}
\caption{The isotopic ratios of the test cases normalized to the abundances obtained with the default model are reported at the end of the C shell for $^{58}$Fe(n,$\gamma$)$^{59}$Fe (upper panel), $^{68}$Zn(n,$\gamma$)$^{69}$Zn (central panel) and $^{85}$Kr(n,$\gamma$)$^{86}$Kr (lower panel).  }
\label{fig:heavy}       
\end{figure}

To discuss the impact of neutron-capture rate of heavy isotopes, in Figure \ref{fig:heavy} we consider the examples of $^{58}$Fe and $^{68}$Zn (both are listed in Table \ref{tab:n_sum}) and $^{85}$Kr at the end of the C-shell. For heavy stable isotopes, we can distinguish two types of propagation impact: one type for the isotopes $^{56-58}$Fe and $^{59}$Co behaving as direct or undirect seeds of the s-process production, and another one for all the other heavier species (including $^{68}$Zn that will be discussed here). 

For the first group, the impact of rate variations may seem quite complicated. By increasing the $^{58}$Fe(n,$\gamma$)$^{59}$Fe rate, in Figure \ref{fig:heavy}, upper panel, the abundance of $^{58}$Fe decreases accordingly. The production of the nearest s-process species is increased as expected, but the heaviest abundances toward the Sr peak decrease. The opposite effect is consistently obtained by reducing the $^{58}$Fe(n,$\gamma$)$^{59}$Fe. For the isotopes in the second group we obtain a more clear propagation. For instance, $^{68}$Zn, which has a low neutron-capture cross section of about 20 mb, is acting as a bottle-neck of the s-process flow: all the heavier s-process products are reduced consistently by decreasing the $^{68}$Zn(n,$\gamma$)$^{69}$Zn (Figure \ref{fig:heavy}, middle panel). 

The unstable isotope $^{85}$Kr is a famous s-process branching point at which both the $^{85}$Kr(n,$\gamma$)$^{86}$Kr and the $^{85}$Kr($\beta$$^{-}$)$^{85}$Rb decay can occur, and its activation is important for both the weak s process and the main s process in AGB stars \cite{pignatari:10, kaeppeler:11, bisterzo:15}. Figure \ref{fig:heavy}, lower panel, shows that its uncertainty is only propagated locally to the isotopes affected by the branching point, i.e., $^{86}$Kr, the Rb isotopes and the s-only isotopes $^{86}$Sr and $^{87}$Sr. Therefore, while the total impact of $^{85}$Kr(n,$\gamma$)$^{86}$Kr is limited (it does not appear in the rate list in Table \ref{tab:n_sum}), robust s-process predictions in the Kr-Sr region need more precise data for this (challenging) reaction rate from the next generation of experimental facilities \cite{reifarth:18}.

\section{Conclusions}
\label{sec:conclusions}

We presented a nuclear sensitivity study for the s process in massive stars. In particular, we focused on the relevant neutron-capture rates between C and Si and between Fe and Zr, for a total of 86 reactions and 173 nucleosynthesis models. A summary table is provided with the twenty neutron-capture rates with the largest impact in the He core and in C-shell conditions. As expected, in the C shell we find a larger impact of light neutron poisons with respect to the He core phase. 

We have made available online the results for all the rates considered, and discussed here four examples: 1) the $^{25}$Mg(n,$\gamma$)$^{26}$Mg impact as representative of the light isotopes, which behave like neutron poisons for the s process; 2) the $^{58}$Fe(n,$\gamma$)$^{59}$Fe case representing the behaviour of the seed Fe and Co nuclei, showing a non-linear propagation to heavier nuclei abundances; 3) the impact of the $^{68}$Zn(n,$\gamma$)$^{69}$Zn rate uncertainty, acting as a bottleneck and the consequent propagation over all the heavier s-process isotopes; 4) the $^{85}$Kr(n,$\gamma$)$^{86}$Kr at the $^{85}$Kr branching-point competing with the $^{85}$Kr($\beta$$^{-}$)$^{85}$Rb decay, with a more localized impact but still relevant for the species involved. 

We also tested the impact of the new library of neutron-capture rates ASTRAL. In the current version (v0.2) ASTRAL includes 122 rates. Compared to the default model, we find changes in the s-process abundances up to a factor of 1.8 (for $^{64}$Zn), with most of the species varying by about 10-30 $\%$.  

Robust weak s-process predictions are a key requirement to define which nucleosynthesis processes contributed to the solar abundances beyond Fe. To achieve this goal, the uncertainty of all relevant neutron-capture reaction rates should be progressively reduced, in particular for rates whose impact is propagated to heavier s-process nuclei. The legacy of Franz K\"{a}ppeler work stands strong in the nuclear astrophysics community and it is paving the way to achieve these goals, as a bright example for future generations of scientists.

\begin{acknowledgements}
The bulk of the work presented here was built by repeating the nuclear sensitivity study designed and recommended by Franz K\"{a}ppeler twenty years ago, which became a crucial driver for several discoveries and scientific publications made with Franz and many other collaborators. We are deeply grateful for Franz's friendship and kindness along these years. MP thanks the support from the NKFI via K-project 138031 and the ERC Consolidator Grant (Hungary) programme (RADIOSTAR, G.A. n. 724560). M.P. and R.R. acknowledge the support to NuGrid from JINA-CEE (NSF Grant PHY-1430152) and STFC (through the University of Hull’s Consolidated Grant ST/R000840/1), and ongoing access to viper, the University of Hull High Performance Computing Facility. M.P. acknowledges the support from the “Lendület-2014” Programme of the Hungarian Academy of Sciences (Hungary). This work was supported by the European Union’s Horizon 2020 research and innovation programme (ChETEC-INFRA – Project no. 101008324), and the IReNA network supported by US NSF AccelNet (Grant No. OISE-1927130). We are deeply grateful for the opportunity to contribute to this volume in memory of Franz K\"{a}ppeler, who has always been a friend and a mentor.

\end{acknowledgements}

\bibliographystyle{spphys}       
\bibliography{file}   

%
%



\end{document}